\documentclass[journal=mamobx,manuscript=article,layout=twocolumn]{achemso}

\usepackage{graphicx}
\usepackage{dcolumn}
\usepackage{float}
\usepackage{bm}
\usepackage{amsmath}
\usepackage{amssymb}

\author{James M. Polson}
\email{jpolson@upei.ca}
\affiliation{Department of Physics, University of Prince Edward Island, 550 University Ave.,
Charlottetown, Prince Edward Island, C1A 4P3, Canada}
\author{Aidan F. Tremblett}
\affiliation{Department of Physics, University of Prince Edward Island, 550 University Ave.,
Charlottetown, Prince Edward Island, C1A 4P3, Canada}
\alsoaffiliation{Current Address: Department of Physics and Physical Oceanography,
Memorial University of Newfoundland, St. John's, NL A1B 3X7, Canada}
\author{Zakary R. N. McLure}
\affiliation{Department of Physics, University of Prince Edward Island, 550 University Ave.,
Charlottetown, Prince Edward Island, C1A 4P3, Canada}

\title{Free energy of a folded polymer under cylindrical confinement}

\begin{document}


\begin{abstract}
Monte Carlo computer simulations are used to study the conformational free energy 
of a folded polymer confined to a long cylindrical tube. The polymer is
modeled as a hard-sphere chain. Its conformational free energy $F$ is
measured as a function of $\lambda$, the end-to-end distance of the polymer.
In the case of a flexible linear polymer, $F(\lambda)$ is a linear function in
the folded regime with a gradient that scales as $f\equiv |dF/d\lambda| \sim N^0 D^{-1.20\pm 0.01}$
for a tube of diameter $D$ and a polymer of length $N$. This is close to the
prediction $f \sim N^0 D^{-1}$ obtained from simple scaling arguments.
The discrepancy is due in part to finite-size effects associated with the de-Gennes
blob model. A similar discrepancy was observed for the folding of a single arm of a 
three-arm star polymer. We also examine backfolding of a semiflexible polymer
of persistence length $P$ in the classic Odijk regime. In the overlap regime,
the derivative scales $f \sim N^0 D^{-1.72\pm 0.02} P^{-0.35\pm 0.01}$, 
which is close to the prediction $f \sim N^0 D^{-5/3} P^{-1/3}$ obtained 
from a scaling argument that treats interactions between deflection segments at the second
virial level. In addition, the measured free energy cost of forming a 
hairpin turn is quantitatively consistent with a recent theoretical calculation.
Finally, we examine the scaling of $F(\lambda)$ for a confined semiflexible chain in 
the presence of an S-loop composed of two hairpins. While the predicted scaling
of the free energy gradient is the same as that for a single hairpin, we observe
a scaling of $f \sim D^{-1.91\pm 0.03} P^{-0.36\pm 0.01}$. Thus, the quantitative 
discrepancy between this measurement and the predicted scaling is somewhat greater for 
S-loops than for single hairpins.
\end{abstract}

\maketitle

\section{Introduction}
\label{sec:intro}

Recent advances in nanofabrication techniques have enabled the systematic of study of the 
physical behavior of single DNA molecules confined to nanochannels.\cite{dai2016polymer,%
reisner2012dna} These studies are largely motivated by such diverse applications as DNA sorting,%
\cite{dorfman2012beyond} DNA denaturation mapping,\cite{reisner2010single,marie2013integrated} 
and genome mapping,\cite{lam2012genome,dorfman2013fluid} each of which exploits the effects of 
confinement on the DNA conformational behavior and dynamics.  The development of such 
applications clearly requires a deep understanding of the physical behavior of polymers 
in nanochannels, and in recent years there has been considerable progress toward this goal. 
Much of this progress is due to the refinement and application of advanced Monte Carlo simulation 
techniques such as the Pruned Enriched Rosenbluth Method (PERM), which enables the simulation 
of very long polymer chains.\cite{hsu2011review,tree2013dna} 
Such studies have been instrumental in characterizing the various 
conformational scaling regimes, which are determined by the degree of confinement in relation 
to the persistence and contour lengths of the polymer.\cite{odijk2008scaling,dai2016polymer} 
Notable among this work is the confirmation of the existence of the extended de~Gennes 
regime\cite{dai2014extended} and the backfolded Odijk regime,\cite{muralidhar2014backfolding} 
which each lie between the classic de~Gennes\cite{deGennes_book} and Odijk 
regimes\cite{odijk1983statistics} upon variation in the confining channel width.

In addition to work characterizing the scaling regimes of polymers confined to nanochannels, 
some effort has been devoted to elucidating the behavior of confined polymers in 
out-of-equilibrium folded states. For example, using fluorescence imaging techniques
Levy {\it et al.}\cite{levy2008entropic} examined the behavior of DNA that was electrophoretically 
driven into a nanochannel in a folded state. They quantified the degree of stretching in the 
overlapping portion of the molecule and monitored the internal dynamics as the
DNA unfolded to its equilibrium linearly ordered state. In a more recent study,
Alizadehheidari {\it et al.}\cite{alizadehheidari2015nanoconfined} examined the
unfolding dynamics of a circular DNA molecule upon transformation to a linear topology by a 
light-induced double-strand break, and in addition compared the equilibrium
conformational statistics of the linear and circular configurations. These experimental
studies have been complemented by molecular dynamics simulation studies of unfolding of 
flexible polymer chains in cylindrical\cite{ibanez2012entropic} and square\cite{ibanez2013hairpin}
nanochannels, where the unfolding time was determined to scale as $DN^2$, where
$D$ is the channel width and $N$ is the polymer length.

The tendency for a nanochannel-confined polymer to unfold arises from the excluded
volume interactions between the portions of the molecule that overlap along the channel.
These interactions tend to stretch the overlapping regions and significantly reduce 
the number of configurations available to the polymer, thus decreasing its conformational
entropy.  The resulting gradient in the free energy with respect to the degree of overlap provides 
the effective force that drives the unfolding. This effect is closely related to the 
segregation of two initially overlapping polymers confined to a narrow channel, a process 
that is also driven by the increase in conformational entropy as polymer overlap decreases.
Such entropy-driven polymer separation is thought to be a factor in the process of
chromosome segregation of replicating bacteria\cite{jun2006entropy,jun2010entropy}
and has been extensively studied using computer simulation methods.\cite{%
jun2006entropy,jun2007confined,arnold2007time,jung2010overlapping,jung2012ring,%
jung2012intrachain,liu2012segregation,dorier2013modelling,racko2013segregation,%
shin2014mixing,minina2014induction,minina2015entropic,chen2015polymer,polson2014polymer}
Another related process is the arm retraction and escape transition for
channel-confined star polymers.\cite{racko2015arm}

\begin{figure}[!ht]
\begin{center}
\includegraphics[width=0.47\textwidth,angle=0]{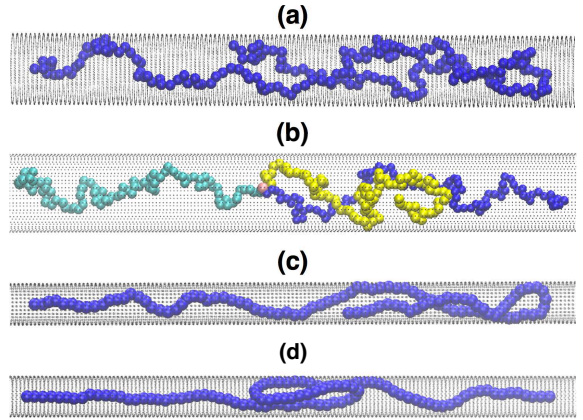}
\end{center}
\caption{
Simulation snapshots illustrating the main systems considered in this study: (a) fully
flexible linear polymer; (b) fully flexible three-arm star polymer; (c) semi-flexible
linear polymer with a hairpin turn; (d) semi-flexible linear polymer with an S-loop.
The images were generated using VMD.\cite{HUMP96}
}
\label{fig:illust}
\end{figure}

Theoretical analyses of unfolding or segregation dynamics obtained from simulations
typically employ analytical approximations using scaling arguments for the conformational 
free energy and its variation with the degree of overlap along the channel. However, 
such approximations are known to suffer from finite-size effects for the system
sizes typically employed in these simulations.\cite{kim2013elasticity} 
For this reason, it is of interest to calculate the free energy functions 
directly and quantify any discrepancy with the theoretical predictions.
Recently, we employed Monte Carlo methods to measure the free energy functions
for segregating polymers in nanotubes and examined the scaling of the functions
with respect to polymer contour length, persistence length, and channel dimensions for
both infinite-length and finite-length tubes.\cite{polson2014polymer} In the present study,
we carry out similar calculations for a single folded polymer confined to a cylindrical
channel. We consider several different variations of the system. First, we examine
a freely jointed polymer chain similar to that employed in previous MD studies of 
unfolding.\cite{ibanez2012entropic,ibanez2013hairpin} We also study a confined three-arm 
star polymer and examine the effect of folding one of the arms on the conformational
free energy. In this case, the results are relevant to the dynamics of arm 
retraction in star polymers, which was the focus of a recent simulation study by
Milchev {\it et al.}\cite{milchev2014arm} In addition to fully-flexible polymers, we
also examine semiflexible chains with hairpin folds. Such systems have been
the subject of much recent study in the context of the backfolded Odijk regime, for which
the persistence length $P$ is of the order of the channel width $D$.\cite{%
muralidhar2014backfolding,muralidhar2016backfolding,muralidhar2016backfolded,werner2016emergence}
In the present case, we consider instead the regime where the condition for the classic
Odijk regime, $P\gg D$, is marginally satisfied and, thus, where the presence of a hairpin 
clearly represents an out-of-equilibrium state. We consider both single hairpins and
S-loops composed of two hairpins. Figure~\ref{fig:illust} shows simulation snapshots of 
the various systems that were examined in this study. The focus in all cases is the 
measurement of the gradient in the free energy with respect to the degree of internal overlap
along the channel, which determines the effective force that drives unfolding. 
The scaling properties of the free energy are compared with the analytical approximations.
While this work is clearly relevant to previous studies in which the free energy
of nanochannel-confined polymers was calculated, to our knowledge it is the first to measure 
the variation of the free energy with the degree of internal overlap of the polymer.

This article is organized as follows. First, we briefly describe the
model employed in the study, following which we outline the MC method used to
calculate the free energy functions. We then present the main results
of the study, which are interpreted and discussed in detail. Results for 
fully flexible linear polymers and star polymers are presented, followed by those for
semiflexible chains in the presence of either a single hairpin fold or an S-loop.
In the final section, we summarize the key findings of this study.

\section{Model}
\label{sec:model}

We employ a minimal model to describe a polymer confined to a cylindrical tube. The
polymer is modeled as a chain of hard spheres, each with diameter $\sigma$.  The pair 
potential for non-bonded monomers is thus $u_{\rm{nb}}(r)=\infty$ for $r\leq\sigma$ 
and $u_{\rm{nb}}(r)=0$ for $r>\sigma$, where $r$ is the distance between the centers of 
the monomers. Pairs of bonded monomers interact with a potential $u_{\rm{b}}(r)= 0$ if 
$0.9\sigma<r<1.1\sigma$ and $u_{\rm{b}}(r)= \infty$, otherwise.  Consequently, the bond 
length fluctuates slightly about its average value. In addition to linear polymers,
we also consider a fully flexible three-arm star polymer with arms of equal length,
each connected to one core monomer. The interactions are all the same as for the linear
polymer case.  The polymer is confined to a 
hard cylindrical tube of diameter $D$. Thus, each monomer interacts with the wall of the 
tube with a potential $u_{\rm w}(r) = 0$ for $r<D$ and $u_{\rm w}(r) = \infty$ for $r>D$,
where $r$ is the distance of a monomer from the central axis of the cylinder. Thus, $D$ 
is defined to be the diameter of the cylindrical volume accessible to the centers of the 
monomers.

Most of the simulations in this study employ fully flexible polymer chains.  However, 
we also consider the effects of bending stiffness for linear polymers. To do this, 
we employ a bending potential associated with each consecutive triplet
of monomers. The potential has the form, $u_{\rm bend}(\theta) = \kappa (1-\cos\theta)$.
The angle $\theta$ is defined at monomer $i$ such that 
$\cos\theta_i \equiv \hat{u}_{i}\cdot\hat{u}_{i+1}$,
where $\hat{u}_i$ is a normalized bond vector pointing from monomer $i-1$ to monomer $i$.
The bending constant $\kappa$ determines the stiffness of the polymer and is related
to the persistence length $P$ by $\exp(-\langle l_{\rm bond}\rangle/P) 
= \coth(\kappa/k_{\rm B}T) - k_{\rm B}T/\kappa$. For our model, 
the mean bond length is $\langle l_{\rm bond} \rangle \approx \sigma$. For the large bending
stiffness considered in this study ($\kappa/k_{\rm B}T \geq 15$), this leads to
$P/\sigma \approx \kappa/k_{\rm B}T - 0.5$.

\section{Methods}
\label{sec:methods}

For the model systems described above, Monte Carlo simulations were 
used to calculate the free energy as a function of the end-to-end distance of the polymer, 
$\lambda$, as measured along the axis of the confining tube. In the case of the star polymer, 
$\lambda$ is defined as the distance of the end of a selected arm to the core monomer. The 
simulations employed the Metropolis algorithm and the self-consistent histogram (SCH) 
method.\cite{frenkel2002understanding} 
The SCH method efficiently calculates the equilibrium probability distribution
${\cal P}(\lambda)$, and thus its corresponding free energy function,
$F(\lambda) = -k_{\rm B}T\ln {\cal P}(\lambda)$.
We have previously used this procedure to measure free energy functions in a
study of polymer segregation,\cite{polson2014polymer} as well as in simulation studies
of polymer translocation.\cite{polson2013simulation,polson2013polymer,polson2014evaluating,%
polson2015polymer}

To implement the SCH method, we carry out many independent simulations, each of which employs a
unique ``window potential'' of the form:
\begin{eqnarray}
{W_i(\lambda)}=\begin{cases} \infty, \hspace{8mm} \lambda<\lambda_i^{\rm min} \cr 0,
\hspace{1cm} \lambda_i^{\rm min}<\lambda<\lambda_i^{\rm max} \cr \infty,
\hspace{8mm} \lambda>\lambda_i^{\rm max} \cr
\end{cases}
\label{eq:winPot}
\end{eqnarray}
where $\lambda_i^{\rm min}$ and $\lambda_i^{\rm max}$ are the limits that define the range
of $\lambda$ for the $i$-th window.  Within each window of $\lambda$, a probability
distribution $p_i(\lambda)$ is calculated in the simulation. The window potential width,
$\Delta \lambda \equiv \lambda_i^{\rm max} - \lambda_i^{\rm min}$, is chosen to be
sufficiently small that the variation in $F$ does not exceed a few $k_{\rm B}T$.
The windows are chosen to overlap with half of the adjacent window, such that
$\lambda^{\rm max}_{i} = \lambda^{\rm min}_{i+2}$.  The window width was typically
$\Delta \lambda = 2\sigma$. The SCH algorithm was employed to reconstruct the unbiased
distribution, ${\cal P}(\lambda)$ from the $p_i(\lambda)$ histograms.  For further
detail of the histogram reconstruction algorithm, see Ref.~\citenum{frenkel2002understanding}.

Polymer configurations were generated by carrying out single-monomer moves using a combination of
translational displacements and crankshaft rotations. Trial moves were accepted with a
probability $p_{\rm acc}$=${\rm min}(1,e^{-\Delta E/k_{\rm B}T})$, where $\Delta E$ is the energy
difference between trial and current states. For simulations of semiflexible chains,
reptation moves were also employed.  Initial polymer configurations were generated 
such that $\lambda$ was within the allowed range for a given window potential. 
Prior to data sampling, the system was equilibrated. As an illustration, for a $N$=500
polymer chain, the system was equilibrated
for typically $\sim 10^7$ MC cycles, following which a production run of $\sim 10^8$
MC cycles was carried out.  On average, during each MC cycle a displacement or rotation
move for each monomer, as well as one reptation move, is attempted once.

In the results presented below, quantities of length are measured in units of $\sigma$ and
energy in units of $k_{\rm B}T$. In addition, the free energy functions obtained from
$F(\lambda)=-k_{\rm B}T\ln {\cal P}(\lambda)$ are shifted such that $F$=0 at the minimum
in all plots of $F$ vs $\lambda$.

\section{Results}
\label{sec:results}

\begin{figure}[!ht]
\begin{center}
\includegraphics[width=0.45\textwidth]{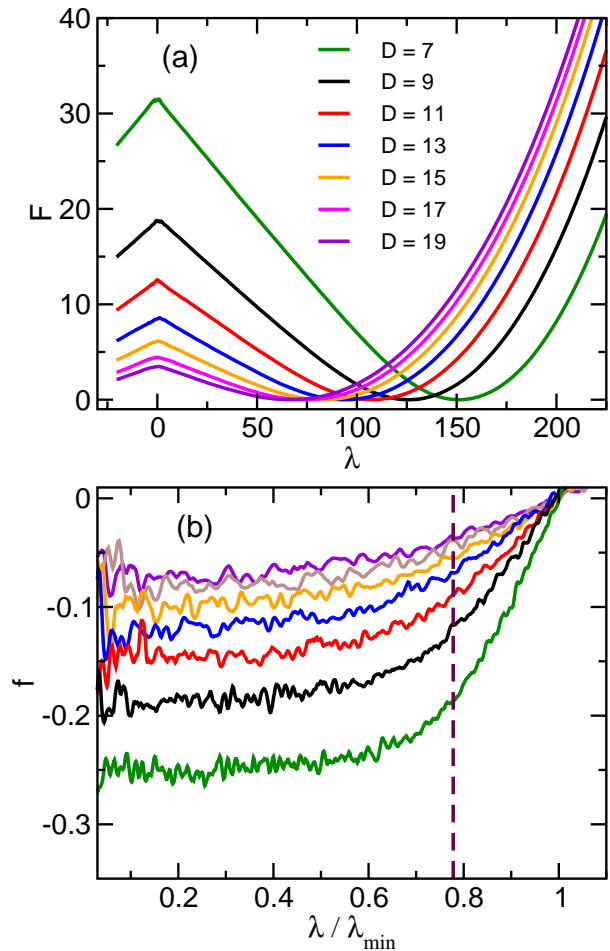}
\end{center}
\caption{
(a) Free energy functions for a flexible polymer of length $N=500$ confined to a cylinder 
of diameter $D$. The curves show the free energy $F$ as a function of the monomer end 
distance $\lambda$ along the cylindrical axis.  Results for several cylindrical diameters 
are shown. 
(b) Derivative of the free energy, $dF/d\lambda$, vs end-monomer separation $\lambda$
as a function of scaled $\lambda$. The results correspond to the data shown in (a). The 
vertical line marks the transition between the folded and non-folded regimes predicted by
the theory of Ref.~\citenum   {milchev2014arm}.  }
\label{fig:F.dFdz.N500}
\end{figure}

Figure \ref{fig:F.dFdz.N500} shows free energy functions for a fully-flexible linear
polymer of length $N$=500 under cylindrical confinement. Results are shown for a 
variety of tube diameters. By symmetry, the curves all satisfy $F(-\lambda)=F(\lambda)$,
though the figure only shows a narrow range of negative $\lambda$.  Each curve has a single
free energy minimum corresponding to the most probable longitudinal end-to-end distance,
which is roughly a measure of the average extension length of the polymer along
the tube. As expected, the location of the free energy minimum, $\lambda_{\rm min}$,
shifts to higher $\lambda$ as $D$ decreases. For $\lambda > \lambda_{\rm min}$, $F$
rises steeply with increasing $\lambda$ due to the reduction in conformational entropy 
associated with stretched conformations. In the regime $\lambda < \lambda_{\rm min}$,
the increase in $F$ with decreasing $\lambda$ gradually becomes linear. This is
evident in Fig.~\ref{fig:F.dFdz.N500}(b), which shows the variation of the derivative 
$dF/d\lambda$ with $\lambda/\lambda_{\rm min}$, calculated using the functions in 
Fig.~\ref{fig:F.dFdz.N500}(a). As $\lambda$ decreases, $dF/d\lambda$ approaches a constant.
The magnitude of $dF/d\lambda$ in this regime increases with deceasing confinement tube 
diameter, $D$, and consequently the height of the free energy barrier,
$\Delta F\equiv F(0)-F(\lambda_{\rm min})$, does as well.
This linear regime corresponds to the case of intramolecular overlap, which is illustrated
in Fig.~\ref{fig:illust}(a). As $\lambda$ decreases and the two end-monomers are brought
closer together, portions of the polymer are forced to overlap. This may occur with a
single backfold, as illustrated in Fig.~\ref{fig:illust}(a), or with two backfolds.
The degree of overlap increases as $\lambda$ decreases, leading to a reduction in
conformational entropy and a corresponding increase in $F$.

The observed trends in the free energy functions can be interpreted using a scaling theory 
developed by Milchev {\it et al.}\cite{milchev2014arm} In that study, an expression was 
derived for the free energy function of a single chain in a tube and the predictions were used
to interpret results for simulations of a confined three-arm star polymer. The theory
employed a free energy function constructed from two terms, one accounting for compression
that uses scaling behavior in the semi-dilute regime, and another term of Pincus form
that accounts for stretching. Using this functional form in cases with and without a
backfold and optimizing the free energy, they predict a transition from a state of uniform 
compression or expansion about the free energy minimum and a backfolded regime to lie at 
$\lambda$=0.779$\lambda_{\rm min}$ (using the notation of the present study). However, 
in Fig~\ref{fig:F.dFdz.N500}(b), we observe instead a gradual transition between the 
backfolded regime the uniformly expanded/compressed regime, with the predicted transition
point (labeled with a vertical dashed line in th figure) only roughly marking the
region of the transition.

\begin{figure}[!ht]
\begin{center}
\vskip 0.1in
\includegraphics[width=0.4\textwidth]{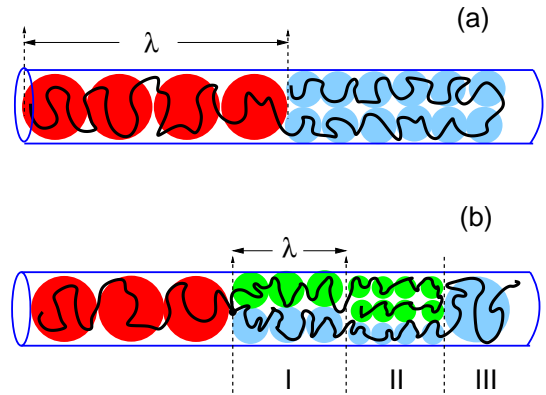}
\end{center}
\caption{Illustration of blob domains for (a) a confined linear polymer with a backfold and
(b) a confined star polymer with one backfolded arm.}
\label{fig:blob_illust}
\end{figure}

Other tests of the accuracy of the theory of Ref.~\citenum{milchev2014arm} can be made using
the simulation results. One such test is the scaling of the entropic force in the backfolded
regime. In this regime, it was shown that the variation of the free energy with end-to-end 
distance is $F(\lambda) = F_0 (2^{1/2\nu} - p \lambda / \lambda_{\rm min})$, where 
$p\approx 0.91$ for a Flory exponent of $\nu\approx 0.588$, and where $F_0\sim ND^{-1/\nu}$ 
is the polymer free energy and $\lambda_{\rm min}\sim ND^{1-1/\nu}$ is the equilibrium extension 
of the polymer, both obtained from free energy minimization with respect to $\lambda$. 
(See Eqs.~(2), (3) and (8) in Ref.~\citenum{milchev2014arm}, and note the change in notation 
to match that used here.) Note that the scaling for $F_0$ and $\lambda_{\rm min}$ is identical
to that predicted from the de~Gennes blob model for the free energy energy and extension
lengths of a polymer in a tube.
From this relation for $F(\lambda)$, it can be easily shown that the entropic force, 
defined as the magnitude of the derivative of the free energy, $f\equiv |dF/d\lambda|$, 
is constant with respect to $\lambda$ and scales as $f \sim N^0 D^{-1}$. 
This prediction also follows from a straightforward application of the de~Gennes
blob picture in conjunction with an approximation for overlapping confined chains that was
suggested in Ref.~\citenum   {jung2012ring}. As illustrated in Fig.~\ref{fig:blob_illust}(a)
the non-overlapping region is approximately of length $\lambda$ and is composed
of $n_{\rm bl}=\lambda/D$ blobs of $g\approx D^{1/\nu}$ monomers. Likewise, in the 
overlapping regime, the monomers may be viewed as two strings of blobs of size $D/\sqrt{2}$,
since they effectively occupy half the cross-sectional area of the tube.\cite{jung2012ring}
Thus, each blob has $g^\prime\approx (D/\sqrt{2})^{1/\nu}$ monomers, and there are 
$(N-(\lambda/D)/g)/g^\prime$ of these smaller blobs present. Noting that each blob contributes
roughly $k_{\rm B}T$ to $F$, it follows that $f \equiv |dF/d\lambda| \sim N^0 D^{-1}$.

\begin{figure}[!ht]
\begin{center}
\includegraphics[width=0.45\textwidth]{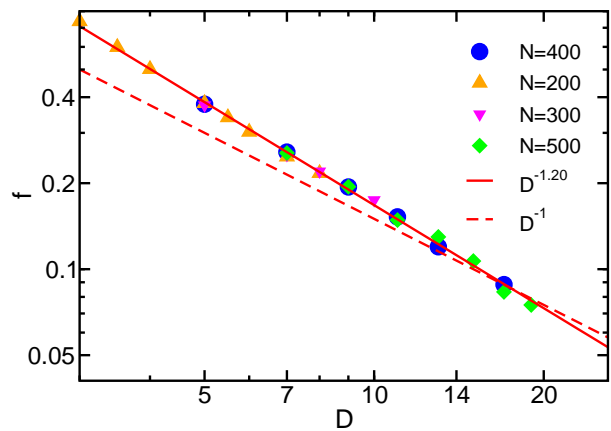}
\end{center}
\caption{
Variation of $f\equiv |dF/d\lambda|$ with the confining cylinder diameter $D$. Results are 
shown for different polymer lengths $N$.  Note that  $f$ is calculated in the 
linear regime of $F(\lambda)$. The solid line shows the power law scaling that yielded
the best fit of the data, i.e. $f \sim D^{-1.20\pm 0.01}$. The dashed line shows
the prediction of the scaling theory of Ref.~\citenum   {milchev2014arm}.
}
\label{fig:dFdz_vs_D}
\end{figure}

Figure~\ref{fig:dFdz_vs_D} shows the variation of $f$ with respect to $D$
calculated from linear fits of the free energy functions in the regime 
$\lambda < 0.5\lambda_{\rm min}$. Results for several chain lengths are shown. 
Consistent with the prediction, there is no significant dependence on $N$,
and the data do satisfy a power law scaling. However, a fit to the data yields
a scaling of $f \sim N^0 D^{-1.20\pm 0.01}$. The fitted curve and a curve for
the predicted $D^{-1}$ scaling are both overlaid on the data in figure, and
the comparison clearly illustrates that the discrepancy of the prediction is significant.
A discrepancy of comparable magnitude was observed for the scaling of the free energy gradient
in our previous study of entropy-driven polymer segregation.\cite{polson2014polymer}
In that case, the scaling of the gradient in the free energy with respect to the 
center-of-mass separation distance was $f\sim N D^{-1.93\pm 0.01}$, compared to the 
predicted scaling of $f \sim N D^{-1.70}$. The physical origins of that discrepancy 
are likely the same as those for the single folded polymer, which will be discussed
below.

\begin{figure}[!ht]
\begin{center}
\includegraphics[width=0.45\textwidth]{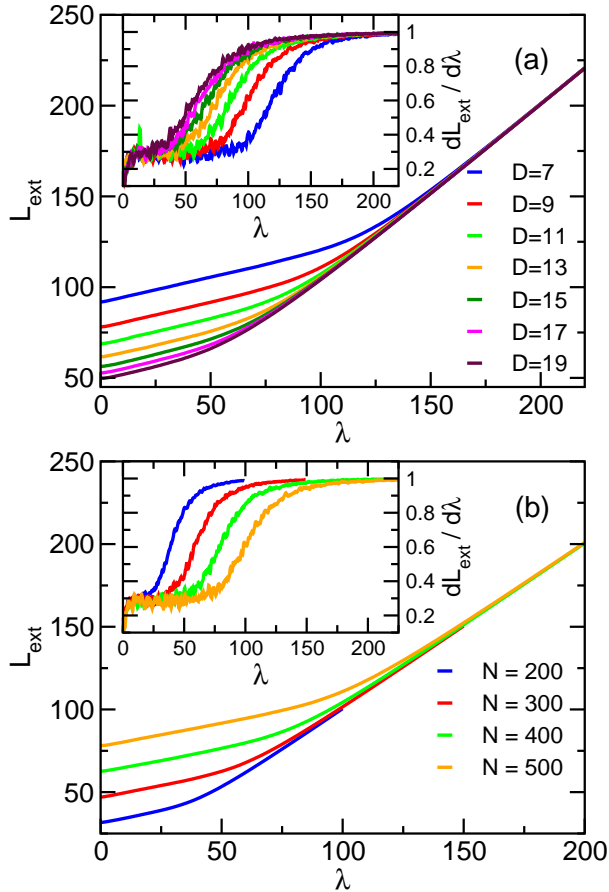}
\end{center}
\caption{
(a) Extension length of the polymer $L_{\rm ext}$ vs end-monomer separation
along the cylinder axis $\lambda$ for a polymer of length $N=500$.
Results for several values of the cylinder diameter $D$ are shown.
The inset shows the derivative $dL_{\rm ext}/d\lambda$ vs $\lambda$.
(b) As in (a), except with fixed $D$=9 and for various $N$.
}
\label{n500_len}
\end{figure}

In addition to the conformational free energy, we also consider the variation
of the mean extension length of the polymer, $L_{\rm ext}$, as a function of
$\lambda$.  Figure~\ref{n500_len}(a) shows results for $N$=500 for various tube 
diameters, while Fig.~\ref{n500_len}(b) shows results for various polymer lengths 
with fixed tube diameter $D$=9. In each graph, the inset shows the derivative 
$dL_{\rm ext}/d\lambda$ vs $\lambda$ for each data set.  As expected, at sufficiently 
high $\lambda$, the polymer is stretched relative to typical equilibrium conformations 
and $L_{\rm ext} \approx \lambda$. This is evident from the overlapping curves for all 
$N$ and $D$ and the limiting behavior of $dL_{\rm ext}/d\lambda\rightarrow 1$ at
high $\lambda$. In the backfold regime at low $\lambda$, $L_{\rm ext}$ is dependent
on both $N$ and $D$. However, independent of $N$ and $D$, the curves in this regime 
are each linear and parallel with one another with a derivative of 
$dL_{\rm ext}/d\lambda\approx 0.27$. The transition between the two regimes occurs
in the vicinity of $\lambda \approx \lambda_{\rm min}$, which lies at lower
$\lambda$ for higher $D$ (as evident in Fig.~\ref{fig:F.dFdz.N500}(a)) and for
lower $N$.

As for the case of the free energy, the behavior of the extension length curves in 
the backfold regime can be understood in the context of the scaling theory of 
Ref.~\citenum   {milchev2014arm}. In that case, it was noted that length $y$ of the 
overlapping portions of the polymer satisfies
$y = 2^{(1-3\nu)/2\nu}(\lambda_{\rm min} - \lambda/0.779)$.
Approximating $L_{\rm ext}\approx \lambda + y$, it follows that 
$dL_{\rm ext}/d\lambda \approx 0.18$, where we use a Flory exponent $\nu\approx 0.588$.
Thus, the observed independence  of $dL_{\rm ext}/d\lambda$ with respect to $N$ and $D$ 
is consistent with the prediction, though the predicted value of the derivative 
is somewhat of an underestimate.

To summarize, the scaling theory captures the qualitative behavior and some
of the quantitative behavior of the free energy and extension length functions.
Unsurprisingly for systems of this size, there are quantitative discrepancies.
Finite-size effects have been studied and quantified in other studies of
polymers confined to narrow channels and suggest that results consistent with
blob model predictions emerge only for diameters of $D>10$.\cite{kim2013elasticity}
In the present case, the extra crowding associated with backfolded polymer
decreases the effective size of $D$ for each overlapping strand, thus potentially
amplifying the finite-size effect. This is also likely a cause of the discrepancy
between observed and predicted scaling of the free energy gradient with respect
to center-of-mass separation for segregating polymers under cylindrical confinement
observed in Ref.~\citenum   {polson2014polymer}. Another possible problem concerns
the validity of the approximation suggested in Ref.~\citenum   {jung2012ring} that
overlapping polymers can each be viewed as strings of blobs of size $D/\sqrt{2}$,
since they effectively occupy half the cross-sectional area of the tube. In recent
work, we have carried out simulations to test this assumption and find significant
quantitative discrepancies from this prediction.\cite{polson2017segregation} Together 
with finite-size effects associated with the de~Gennes blob model, this factor likely 
contributes to the observed inconsistency.

There is a third possible source for the discrepancy, for which we now provide evidence 
to show is not significant.  It concerns a point noted in Ref.~\citenum   {milchev2014arm} 
regarding the possible presence of two backfolds. The calculated free energy
of the polymer depends on the end-to-end distance, which determines the degree
of polymer overlap. However, configurations with two backfolds with the same
$\lambda$ can produce the same degree of polymer overlap, and it was suggested
that these additional configurations lead to logarithmic corrections to the
prediction for free energy. It is possible that this effect may contribute
to the discrepancies between our measurements and the predictions. To test this
possibility, we have carried out a modified simulation that imposes a single
backfold (at most) on the system for any given $\lambda$. To do this, we
insert a hard wall cap in the cylinder, at the center of which a single monomer is fixed.
During the simulations, we carry out reptation moves in addition to the regular MC trial moves
so that the sequence position of the monomer fixed to the cap, $m$, can take
a value from 1 to $N$. Employing the SCH method using $m$ as the fluctuating
variable, we measure the free energy function $F(m)$. Clearly, there can only
be a single backfold, located at the position of monomer $m$. Furthermore, a given value 
of $m$ corresponds to a unique value of the degree of internal polymer overlap along
the tube. If the results obtained using this model are closer to predictions
from the scaling theory, then this would provide evidence that the previous discrepancies
were partially due to the presence of multiple backfolds. In addition to 
helping resolve this question, this calculation has the side benefit 
of clarifying a particular issue in the context of the related process of polymer
translocation. Specifically, when a polymer translocates through a nanopore
in a barrier, the likelihood of initially capturing the polymer from one of
its ends or at another point along the contour in a folded configuration
is expected to be affected by the degree of polymer confinement prior to
translocation. The present calculation provides a means to quantify the likelihood
for a given capture position along the chain i.e. higher $F(m)$ will correspond
to a lower likelihood of capture at monomer $m$.\cite{ding2016flow}

\begin{figure}[!ht]
\begin{center}
\includegraphics[width=0.45\textwidth]{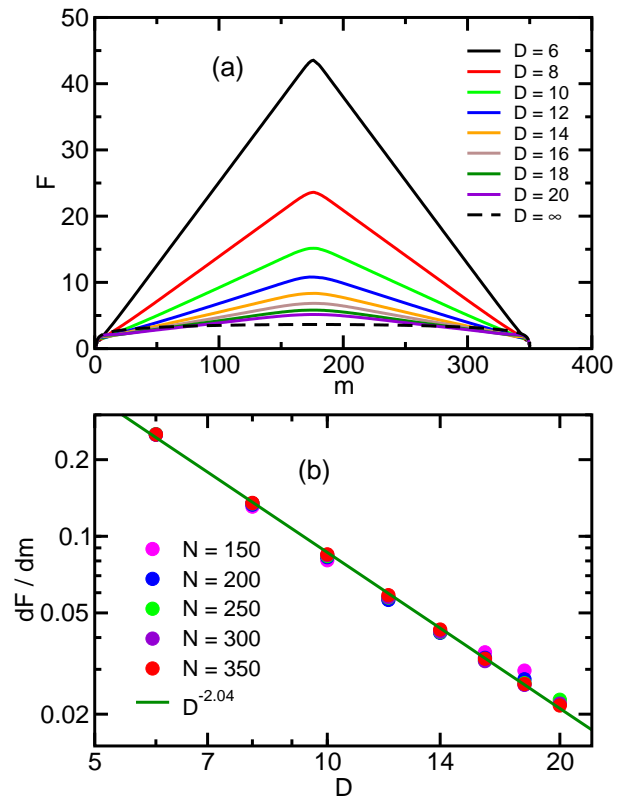}
\end{center}
\caption{
(a) Free energy vs monomer index $m$ for a flexible polymer confined to a cylinder 
of diameter $D$ and semi-infinite length. Monomer $m$ is fixed to a hard cap in the cylinder. 
Results are shown for a polymer of length $N$=350 for several different values of $D$.
(b) Derivative $dF/dm$ vs tube diameter $D$. The derivative was calculated from a fit
to the free energy functions in (a) in linear regime, as well as data for other
polymer lengths. The solid line is a fit of the $N$=350 data to a power-law function, 
which yielded a scaling of $dF/dm \sim D^{-2.04\pm 0.02}$.  }
\label{fig:F_zak.N350}
\end{figure}

Figure~\ref{fig:F_zak.N350}(a) shows the variation in the free energy with the
index of the monomer that is tethered to the cap in the semi-infinite cylinder.
Results for a polymer length of $N$=350 and several different tube diameters
are shown. As expected, $F$ is greatest when $m$=$N/2$, in which case the 
polymer subchains on either side of the anchored monomer are of equal contour length,
leading to maximum overlap along the tube. (Note that the extension lengths of
the subchains along the channel will differ slightly in this case, as noted in 
Ref.~\citenum{milchev2014arm}, and thus a little fewer than $N/2$ monomers of the longer
subchain will overlap with those of the other. This leads to a slightly lower
free energy maximum at $m$=$N/2$ than would otherwise be the case.)
As the $m$ increases or decreases
from this point, the subchain lengths become increasingly unequal, and the shorter
subchain overlaps only partially with the longer subchain. In turn, there is less
crowding for a greater number of monomers, leading to a decrease in the
free energy.  For sufficiently small $D$, the variation is linear over most of
each half of the curve, though the functions tend to be slightly less linear with 
increasing $D$.

To understand this effect, the blob model can once again be used. For $m<N/2$,
the shorter subchain is of length $m$, while the longer subchain is of length
$N-m$. Clearly, the portion of the subchain that overlaps the short subchain
is also of length $m$, and thus the remaining (i.e. non-overlapping) portion 
of the longer subchain is of length $N-2m$. The confined polymer system then can
be viewed as two chains of length $m$ and one of length $N-2m$ with a corresponding
free energy of
$F(m)/k_{\rm B}T \sim 2 m\left(D/\sqrt{2}\right)^{-1/\nu} + (N-2m) D^{-1/\nu}$,
from which it follows, $dF/dm\propto N^0D^{-1/\nu} = N^0D^{-1.70}$. By symmetry, 
the derivative is equal in magnitude and opposite in sign for $m>N/2$. 
Thus, the decrease in the derivative (and, therefore, the free energy barrier height)
is qualitatively consistent with this prediction. Figure~\ref{fig:F_zak.N350}(b)
shows the variation in $dF/dm$ with $D$ for several different polymer lengths.
Consistent with the prediction, there is no dependence on $N$ except for small
finite-size variations at large $D$. However, a fit to the data for $N$=350
yields a scaling of $dF/dm \propto D^{-2.04\pm 0.02}$, which represents a significant
discrepancy from the predicted scaling. The magnitude of this discrepancy
is essentially the same as between the predicted and observed scaling of the data in 
Fig.~\ref{fig:dFdz_vs_D}. We conclude that presence of two backfolds was not
the main source of this disagreement.

\begin{figure}[!ht]
\begin{center}
\includegraphics[width=0.45\textwidth]{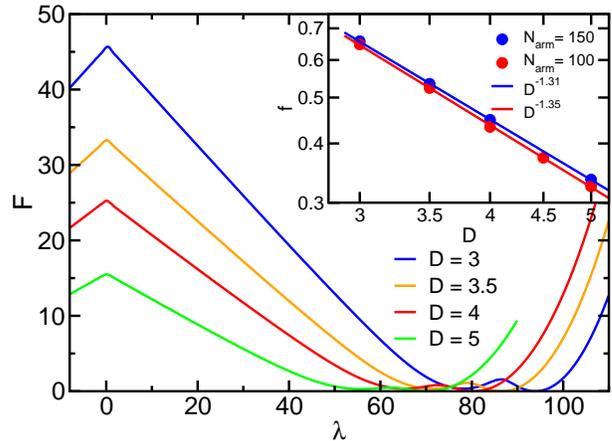}
\end{center}
\caption{
Free energy $F$ vs distance $\lambda$ between an end-monomer and the core monomer along the
cylinder axis $\lambda$ for a three-arm star polymer. Results are shown for
an arm length $N_{\rm arm}=150$ for several values of the cylinder diameter $D$. 
The inset shows the derivative $f\equiv |dF/d\lambda|$ in the linear domain at low $\lambda$
vs tube diameter $D$ for systems with $N_{\rm arm}$=150 and $N_{\rm arm}$=100.
The solid lines are power-law fits to each data set, which yielded a scaling
of $f \sim D^{-1.35\pm 0.01}$ for $N_{\rm arm}$=100 and $f \sim D^{-1.31\pm 0.01}$ for 
$N_{\rm arm}$=150.
}
\label{F_star_N150}
\end{figure}

We now examine the behavior of a confined star polymer.
Figure~\ref{F_star_N150} shows free energy functions for a three-arm star polymer 
of arm-length $N_{\rm arm}$=150 for several different values of $D$. In this case 
$\lambda$ is the longitudinal distance between the core monomer and the end-monomer
of a single chosen arm. We consider values of $N_{\sf arm}$ and $D$ with a sufficiently
high free energy barrier that neither of the other arms has an appreciable chance of
crossing while the chosen arm is brought to the selected range of $\lambda$ for each
simulation used to calculate $F(\lambda)$.  The curves are qualitatively similar to those of 
Fig.~\ref{fig:F.dFdz.N500}(a). As before, there is a linear regime at sufficiently low
$\lambda$, in which the arm that defines $\lambda$ is backfolded.  One notable difference 
from the results for the linear polymer is the presence of two free energy
minima separated by a small barrier. This feature has an origin similar to the behavior
observed in Ref.~\citenum   {milchev2014arm} for two overlapping polymers confined to a 
cylinder and both tethered to a cap at one end. In that case, the lowest free energy
state corresponds to that of slightly different elongation lengths for the polymers,
with a small barrier located at equal extension length separating states where the
roles of compressed and elongated polymers are exchanged. In the present case, the two 
arms of the star polymer connected to the core monomer play the same role, giving rise to
the same effect.

The free energy gradient $f$ for the linear portion of the free energy function
is plotted as a function of $D$ in the inset of the figure. Results are shown for
arm lengths of $N_{\rm arm}$=100 and 150. There is a slight dependence of
$f$ on the arm length, with slightly larger values of $f$ for the greater $N_{\rm arm}$.
Fitting the results to a power law yields a scaling of $f \sim D^{-1.35\pm 0.01}$
for $N_{\rm arm}$=100 and $f \sim D^{-1.31\pm 0.01}$ for $N_{\rm arm}$=150.

In principle, these results can be understood using a simple scaling analysis employing 
the de~Gennes blob model. Consider the case where two arms lie on one side of the
core monomer, one of which is the arm selected to define $\lambda$. If this arm
has a backfold (while the other does not), then there are three distinct regimes 
for these two arms along the cylinder, which are labeled in Fig.~\ref{fig:blob_illust}(b).
In regime~I, the overlapping polymer arms can be viewed as two strings of blobs,
each effectively occupying half the cross sectional area of the tube. Following
the approach taken in Ref.~\citenum   {jung2012ring}, the blobs can be viewed
as being confined to a tube of effective diameter $D/\sqrt{2}$. The span of this
regime is approximately $\lambda$. Likewise, in regime~II where the backfolded chain 
lies, the polymer subchains can be viewed as three strings of blobs of effective 
diameter $D/\sqrt{3}$. Finally, in regime~III, there is only one string of blobs
for the stretched arm, each of diameter $D$. Varying $\lambda$ will change the
number of blobs in each regime, each of which contributes on the order of $k_{\rm B}T$
to the free energy. Varying $\lambda$ does not affect the number of blobs for the
arm on the other side of the core. Following the same approach used earlier to
derive the scaling for the backfolded linear polymer, it can easily be shown
that the predicted scaling is $f \sim N_{\rm arm}^0 D^{-1}$. The poor agreement
with the measured scaling is not surprising here. In part, it is due to the very
small values of $D$ used here, for which the blob model is expected to be inaccurate.
In addition and as noted above, the approximation of Ref.~\citenum   {jung2012ring} employed here
has also been shown to suffer quantitative inaccuracies.\cite{polson2017segregation}
It is expected that the scaling will improve for larger values of $D$ and $N_{\rm arm}$.
However, such calculations are not computationally feasible for us at present.

\begin{figure}[!ht]
\begin{center}
\includegraphics[width=0.45\textwidth]{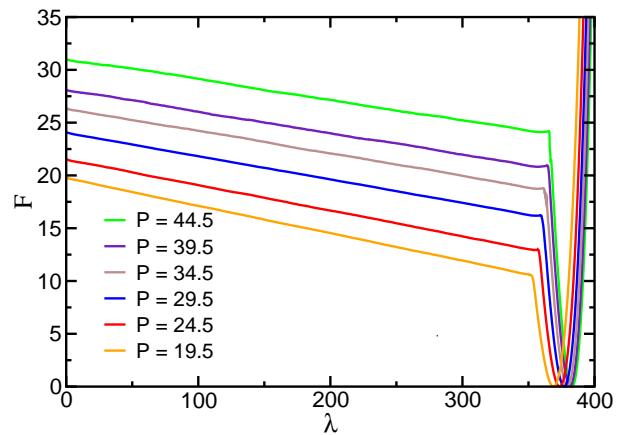}
\end{center}
\caption{
Free energy vs $\lambda$ for a semiflexible linear polymer of length $N$=400
in a cylindrical tube of diameter $D$=6. Results for several values of the persistence 
length $P$ are shown. }
\label{F.N400.D6}
\end{figure}

Let us now turn to the folding behavior of longitudinally confined semiflexible polymers 
in the classic Odijk regime. Figure~\ref{F.N400.D6} shows free energy functions for 
semiflexible polymers of length $N$=400 in a tube of diameter $D$=6. Results are shown 
for several different values of persistence length $P$ with $P/D \geq 3$, thus marginally 
satisfying the conditions for the Odijk regime. The functions have a form that
is qualitatively different from those for the flexible chains shown in 
Fig.~\ref{fig:F.dFdz.N500}(a). In this case there are two distinct regimes, one
corresponding to a deep free energy well at high $\lambda$, and a second with a
linear variation of $F$ with $\lambda$ over a broad range of $\lambda$ to the
left of the well, with a sharp boundary between the two. Similar results were
observed for other values of $N$ (data not shown). As the chains stiffen, the position 
of the free energy minimum shifts slightly to higher $\lambda$, and its width narrows. 
Thus, the polymers become more elongated and the fluctuations in the extension length
decrease, in accord with established results for confined semiflexible chains in the
classic Odijk regime.\cite{odijk1983statistics} 
The increase in $F$ as $\lambda$ decreases from its value at the minimum
corresponds physically to the formation of a hairpin turn. The free energy cost of
forming the hairpin is dominated by the bending energy of the chain, but there is 
also a significant entropic contribution as well.\cite{odijk2006dna,chen2017conformational} 
The more gradual change in $F$ in the linear regime corresponds to the change in the degree
of overlap of the stiff subchains. As $\lambda$ decreases, the overlap increases
leading to a reduction in conformational entropy, thus increasing $F$.

\begin{figure}[!ht]
\begin{center}
\includegraphics[width=0.45\textwidth]{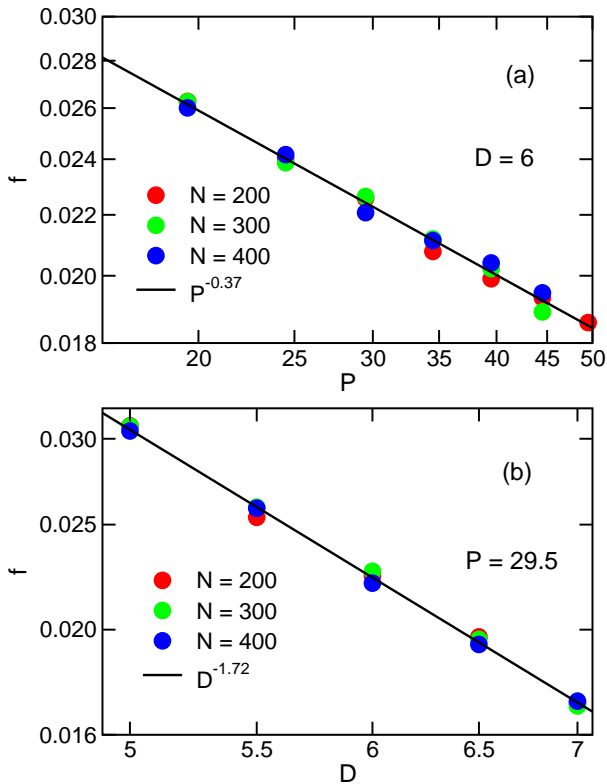}
\end{center}
\caption{
(a) Free energy gradient $f \equiv |dF/d\lambda|$ in the linear overlap regime vs persistence 
length $P$ for a semi-flexible linear polymer in a cylinder of diameter $D$=6. Results for
several polymer lengths are shown. The solid line is a fit to all of the data.
(b) $f$ in the linear overlap regime vs tube diameter for a semi-flexible
polymer of persistence length $P$=29.5. The solid line is a fit to all of the data.}
\label{dFdk.P.D6}
\end{figure}

Figure~\ref{dFdk.P.D6}(a) shows $f$ vs $P$ for a fixed tube diameter of $D$=6, while 
Fig.~\ref{dFdk.P.D6}(b) shows $f$ vs $D$ for a fixed persistence length of $P$=29.5. In 
each case, results for several polymer lengths are shown. No clear dependence on
$N$ is evident. The derivatives were calculated
from fits to curves such as those in Fig.~\ref{F.N400.D6} for the linear regime, where
a single hairpin turn was present and there was partial intrapolymer overlap. Fits
to the data suggest a scaling of $f \sim N^0 P^{-\alpha} D^{-\beta}$, where
$\alpha=0.37\pm 0.01$ and $\beta=1.72\pm 0.02$.

A simple explanation for the scaling behavior of $f$ follows from using
arguments similar to those employed by Odijk in a theoretical study of backfolding 
regimes of DNA in nanochannels.\cite{odijk2008scaling} Recall that the confinement 
free energy of a single semiflexible chain in this regime is given by 
$F_1(L) \sim k_{\rm B}T L/l_{\rm def}$, where $L$ is the contour length of the polymer and
the total Odijk deflection length satisfies $l_{\rm def} \approx D^{2/3}P^{1/3}$. 
The confinement free energy of a backfolded chain can be approximated
$F_{\rm c} = F_1(L-h) + F_{\rm h} + F_{\rm int}$, where the $h$ is the length
of the hairpin fold, $F_1(L-h)$ is the Odijk free energy for the two subchains
outside the fold, $F_{\rm int}$ is the interaction between the strands and
$F_{\rm h}$ is the free energy of the hairpin. In the regime of interest,
only $F_{\rm int}$ depends on $\lambda$, so we neglect the other two terms.
To estimate $F_{\rm int}$, we use an approximation for the free energy of a system
of long, hard rigid rods. For that system, the interaction free energy 
in the 2nd virial approximation is given by $F^{\rm (int)}/k_{\rm B}T = (l^2\sigma N^2/V)
\langle |\sin\gamma|\rangle$ for $N$ rods of length $l$ and diameter $\sigma$ confined
to a volume $V$, where $\gamma$ is the angle between the rods.\cite{odijk1986theory}
When the rods are highly aligned, $\langle|\sin\gamma|\rangle \sim \sqrt{\langle\theta^2\rangle}$,
where $\theta$ is the angle between the rod and the alignment direction. To apply these
results to the present system, we treat each deflection length $l_{\rm d}$ as a rigid 
rod and substitute $l\rightarrow l_{\rm d}\sim D^{2/3}P^{1/3}$. In addition,
we assume that the alignment arises principally from confinement, such that
$\langle\theta^2\rangle \sim (D/l_{\rm d})^2$.  Further, we note
that $V$ is the volume over which the intermolecular segments overlap. This is given
by $V\sim l_{\rm ov} D^2$, where the overlap region is $l_{\rm ov} \approx (L-h-\lambda)/2$.
Finally, replacing $N$ with the number of deflection lengths of the two overlapping segments,
$N\rightarrow 2l_{\rm ov}/l_{\rm def}$, it can be shown that 
$F\sim (L-h-\lambda)D^{-5/3}P^{-1/3}$ plus terms independent of $\lambda$. 
Thus, we predict that $f\equiv |dF/d\lambda| \sim N^0 D^{-\beta}P^{-\alpha}$,
where $\beta=5/3$ and $\alpha=1/3$. These predictions are close to the measured values
of $\beta=1.72\pm 0.02$ and $\alpha=0.37\pm 0.01$. The small quantitative discrepancy
may arise from the fact that the system only just marginally satisfies the condition defining 
the Odijk regime that $P\gg D$, as well as from other approximations that have been employed.
By comparison, in our related previous study of the segregation of overlapping confined 
semiflexible chains, the scaling exponents for the free energy gradient were measured to be 
$\beta\approx 2$ and $\alpha\approx 0.37$.\cite{polson2014polymer} The larger discrepancy 
for the scaling with respect to $D$ in that study is likely due to the narrower tube 
diameters considered, i.e.  $2\leq D \leq 5$, in contrast to the range here of $5\leq D \leq 7$. 
In the present study, the wider tubes correspond to lower packing fraction, for which the 
second virial approximation is better suited to describe interactions between the chain segments. 

\begin{figure}[!ht]
\begin{center}
\includegraphics[width=0.45\textwidth]{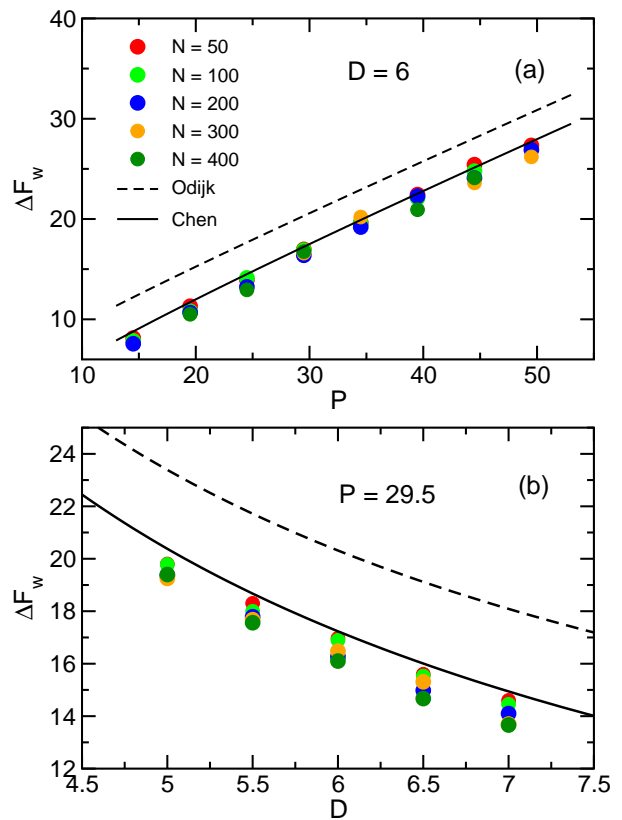}
\end{center}
\caption{(a) Free energy well depth $\Delta F_{\rm w}$ vs persistence length $P$ for a tube 
diameter of $D$=6. Results for various polymer lengths are shown. 
The dashed and solid lines are the hairpin turn free energy values predicted by Odijk in 
Ref.~\citenum   {odijk2006dna} and by Chen in Ref.~\citenum   {chen2017conformational},
respectively. (b) As in (a), except $\Delta F_{\rm w}$ vs tube diameter $D$ for persistence 
length $P$=29.5.
}
\label{fig:delFw}
\end{figure}

We now consider the deep wells in the free energy functions. As noted earlier, the wells 
are associated with the formation of hairpin turns of confined semiflexible polymers.
Figure~\ref{fig:delFw} shows the dependence of the free energy well depth on the
polymer persistence length for various polymer lengths. As expected, there is no
significant dependence of $\Delta F_{\rm w}$ on the polymer length, since the 
well depth is a measure of the free energy required to form a hairpin turn.
Over the range of $P$ considered here, we note that $\Delta F_{\rm w}$ increases 
roughly linearly with $P$.  Overlaid on the data are two theoretical predictions. 
The first is that for a mechanical model developed by Odijk.\cite{odijk2006dna,odijk2008scaling} 
In that work, it was noted that the effects of entropic depletion on the hairpin
make an appreciable contribution to the hairpin free energy in addition
to the bending energy. A more recent and far more rigorous theoretical analysis by 
Chen\cite{chen2017conformational} used the Green's function formalism to determine the 
hairpin free energy and the global persistence length. Results obtained using the latter 
were found to be consistent with the simulation results of Muralidhar and
Dorfman in their study of the backfolded Odijk regime.\cite{muralidhar2016backfolding} 
An analytical representation to Chen's
numerical solution for the hairpin free energy, $F_{\rm hp}$, is shown in the figure.
It is qualitatively similar to Odijk's prediction, though with somewhat lower values of 
$\Delta F_{\rm w}$. This arises from the failure of Odijk's theory to account for
the orientational entropy of the hairpin planes.\cite{chen2017conformational}
(Note that the calculations for the Odijk predictions were carried out using a
correction to an error that was noted by Chen.\cite{chen2017conformational})
As evident in the figure, the predictions of Chen are much closer to the simulation
data than that of Odijk. There appears to be a very slight overestimate of $\Delta F_{\rm w}$
by an amount $\lesssim k_{\rm B}T$. This effect may be due to discretization of the polymer
and may also be associated with the small fluctuating bond length used in our model, neither of
which are present in the theoretical model employed in Ref.~\citenum{chen2017conformational}. 

\begin{figure}[!ht]
\begin{center}
\includegraphics[width=0.45\textwidth]{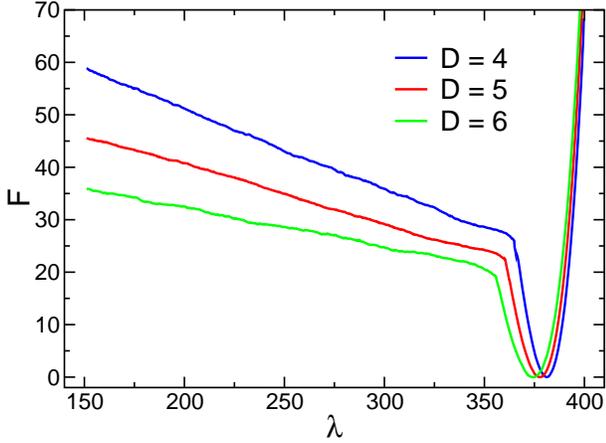}
\end{center}
\caption{
Free energy vs $\lambda$ for a semi-flexible cylindrically confined linear polymer 
with an S-loop. Results are shown for a polymer of length $N$=400,  
persistence length $P$=24.5 and for several values of the tube diameter $D$. }
\label{F_N400_Sloop}
\end{figure}

We turn finally to a brief look at the behavior of S-loop configurations in nanotubes.
The structure of such a formation is illustrated in Fig.~\ref{fig:illust}(d).
In order to prevent the formation of single hairpin turns, the polymer ends are
tethered perpendicular to the centers of virtual caps. Note that this has the effect
of significantly reducing the conformational freedom of the polymer near the ends,
which in turn will affect results, especially for short polymers.
Figure~\ref{F_N400_Sloop} shows free energy functions for a linear semiflexible
polymer chain with such constraints imposed its the ends. Since single hairpins are 
precluded, the polymer forms an S-loop composed of two hairpin turns as the end-to-end 
distance $\lambda$ decreases. The functions
are qualitatively similar to those for polymers with a single hairpin turn in
Fig.~\ref{F.N400.D6}. In this case, the linear regime at lower $\lambda$ 
corresponds to end separations where the S-loop is present. As $\lambda$
decreases and the ends approach each other, the overlap region of the loop
increases and conformational entropy decreases, leading to an increase in $F$.
The depth of the free energy well located at high $\lambda$ corresponds roughly to the
free energy cost of forming the S-loop.

\begin{figure}[!ht]
\begin{center}
\vskip 0.1in
\includegraphics[width=0.45\textwidth]{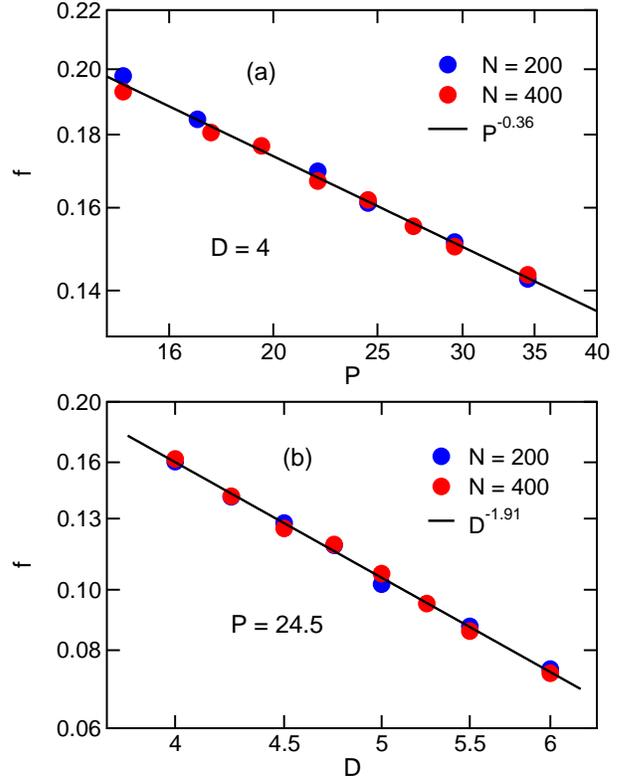}
\end{center}
\caption{
(a) $f\equiv |dF/d\lambda|$ in the S-loop regime vs persistence length $P$ for a 
cylindrically confined semi-flexible linear polymer. Results are shown for a confinement
tube of diameter $D$=4 and polymer lengths of $N$=200 and $N$=400. The solid line is a fit
to both data sets.
(b) $f$ in the S-loop regime vs confinement tube diameter $D$ for a
semi-flexible linear polymer of persistence length $P$=24.5. Results for
$N$=200 and $N$=400 are shown. The solid line is a fit to both data sets.}
\label{dFdkappa.P.R2.5}
\end{figure}

Figure~\ref{dFdkappa.P.R2.5}(a) shows the variation of $f$ with $P$ 
in the linear regime for $D$=4, while Fig.~\ref{dFdkappa.P.R2.5}(b) shows $f$ 
with $D$ for $P$=24.5. In each case, results for $N$=200 and $N$=400 are shown.
As in the case of the results for a single hairpin turn, there is no significant
dependence on the polymer length.
A fit to all of the data suggests a scaling of $f \sim D^{-\beta} P^{-\alpha}$,
where $\alpha$=$1.91\pm 0.03$  and $\beta$=$0.36\pm 0.01$. 

To interpret these scaling results, we employ the same theoretical approach as
that used for the single-hairpin data. In this case, we note that the overlap volume
$V\sim l_{\rm ov}D^2$ has an overlap length $l_{\rm ov}\approx (L-2h-\lambda)/2$,
where $L$ is the polymer contour length and $h$ is the length of each hairpin.
In addition, since there are three polymer segments in the overlap region rather
than two in the case of the single hairpin, there are $3l_{\rm ov}/l_{\rm def}$
deflection lengths in the overlap region, in contrast to $2l_{\rm ov}/l_{\rm def}$
for the single hairpin. Otherwise, the analysis is identical, leading to the same
predicted scaling exponents of $\alpha$=1/3 and $\beta$=5/3. In addition, we
expect that the ratio of the derivatives for the S-loop and single hairpin
is $f_{\rm S}/f_{\rm hp}=(3/2)^2=2.25$. The predicted scaling exponents 
are somewhat less accurate for the S-loop that for the single-hairpin. In addition, 
we find, for example, that $f_{\rm S}/f_{\rm hp}\approx 3.1$ at $D$=6 
and $P$=24.5, which is somewhat larger
than the predicted value. The discrepancies likely arise from multiple sources. We note
that for fixed $D$, the density of deflection segments is larger for the S-loop,
perhaps leading to a reduction in the accuracy of the 2nd virial approximation
to describe excluded volume interactions. Further, as noted above, the reduction
in orientational freedom of the ends tethered to the virtual end caps likely affect
the results. Additional simulations using much longer chains should clarify this matter,
but such calculations are not currently feasible.

\section{Conclusions}
\label{sec:conclusions}

In this study, we have used Monte Carlo simulations to study the conformational free
energy $F$ of folded polymers in cylindrical channels. Specifically, the simulations were 
used to measure the variation of $F$ with respect to the end-to-end distance $\lambda$, 
which determines the degree of internal overlap along the tube. The gradient in 
the free energy, $f\equiv |dF/d\lambda|$, is the effective force that drives unfolding toward 
the polymer's equilibrium state and thus is essential for a meaningful analysis of unfolding 
kinetics measurements in experiments and simulations. The main goal of this work was to 
compare the scaling properties of $f$ with predictions from scaling arguments.
For fully-flexible chains, we find that the gradient scaling of $f\sim N^0 D^{-1.20\pm 0.01}$,
for chains of up to $N$=500 monomers and cylinder diameters of $D$=$3-18$, in units of monomer
diameter. This differs appreciably from the prediction $f\sim N^0 D^{-1}$, a discrepancy that
is likely due to finite-size effects and deficiencies in the approximation used to account
for overlap, i.e. that overlapping polymer strands behave as noninteracting polymers in
effective tubes of size $D/\sqrt{2}$.\cite{jung2012ring} 
A similar and even greater discrepancy was observed 
for the folding free energy for a single arm of a star polymer. The transition from a 
uniformly compressed linearly ordered polymer to a backfolded polymer was found to be gradual, 
in contrast to a recent theoretical prediction. In the case of confined semiflexible polymers
in the classic Odijk regime, we find a free energy gradient scaling of
$f\sim N^0 D^{-1.72\pm 0.02} P^{-0.37\pm 0.01}$, which is close to the 
prediction $f\sim N^0 D^{-5/3} P^{-1/3}$ obtained by treating interactions between
deflection segments at the 2nd virial level. In the case of S-loops, the agreement
was somewhat poorer, perhaps due to the higher segment density in the overlap regime and
a resulting breakdown in the 2nd virial approximation. Finally, we note that the measured
free energy of a hairpin turn was quantitatively consistent with a recent theoretical
prediction.\cite{chen2017conformational}

The work carried out in this study can be extended in various directions.
One straightforward and useful extension is an examination of other channel shapes.
Although the conformational behavior of polymers confined to cylindrical channels 
has been the subject of a number of theoretical studies, understanding the effects of 
confinement in {\it rectangular} channels tends to be more directly relevant to experimental 
studies.  Recent studies employing square or rectangular channels using computer 
simulation\cite{muralidhar2016backfolding,muralidhar2016backfolded} 
and scaling theories\cite{werner2015scaling} 
have highlighted the importance of the channel shape and size
on polymer scaling regimes. In the future, it will be useful to employ
the methods of the present study to characterize the free energy functions for
folded polymers in such channels. Among other points of interest, it is expected that the 
entropic contributions to the hairpin free energy of semiflexible polymers will 
differ appreciably from that for cylindrical channels. Another useful extension
would be an examination the free energy functions in the case of $P\sim D$.
Here, the calculations could be used to quantify the degree of validity of the 
assumed form of the interaction between overlapping polymer subchains employed in the 
theory of the backfolded Odijk regime, as well as extend range of verification
of Chen's prediction of the hairpin free energy. Finally, it would be useful
to carry out molecular dynamics simulations to examine the ability of employing the 
entropic force obtained from the free energy functions to accurately describe the 
kinetics of unfolding.

\begin{acknowledgement}
This work was supported by the Natural Sciences and Engineering Research Council of Canada (NSERC)
Discovery Grants Program.  We are grateful to the Atlantic Computational 
Excellence Network (ACEnet) and WestGrid for use of their computational resources.
\end{acknowledgement}


\providecommand{\latin}[1]{#1}
\makeatletter
\providecommand{\doi}
  {\begingroup\let\do\@makeother\dospecials
  \catcode`\{=1 \catcode`\}=2 \doi@aux}
\providecommand{\doi@aux}[1]{\endgroup\texttt{#1}}
\makeatother
\providecommand*\mcitethebibliography{\thebibliography}
\csname @ifundefined\endcsname{endmcitethebibliography}
  {\let\endmcitethebibliography\endthebibliography}{}

\end{document}